\documentclass[12pt]{article}
\usepackage{pic03}
\usepackage{hyperref}
\usepackage{url}
\usepackage{graphicx}

\begin{document}

\title{\bf SOLAR MODELS AND SOLAR NEUTRINOS}

\author{John N. Bahcall \\
{\em School of Natural Sciences,
Institute for Advanced Study,
 Princeton, NJ 08540, USA}}
\maketitle
%
%
%
%
\vspace{4.5cm}
%

\baselineskip=14.5pt
\begin{abstract}
I summarize $40$ years of development of the standard solar model that is used to predict
solar neutrino fluxes and then describe the current uncertainties in the predictions. I will also attempt
to explain why it took so long, about three and a half decades, to reach a consensus view that new
physics is being learned from solar neutrino experiments.
\end{abstract}
\newpage

\baselineskip=17pt

\section{Introduction}
\label{sec:introduction}

I begin in Section~2,  with a tribute to Ray Davis and Bruno Pontecorvo. In Section~3, I present a
concise history of the development of the standard solar model that is used today to predict solar
neutrino fluxes.  I describe in Section~4 the currently-estimated uncertainties in the solar neutrino
predictions\footnote{Where contemporary numbers are required in this review, I use the results from the
BP00 solar model, ApJ 555 (2001) 990, astro-ph/0010346.}, a critical issue for existing and future solar
neutrino experiments. I also present a formula that gives the ratio of the rates of the $^3$He-$^3$He and
the $^3$He-$^4$He reactions as a function of the p-p and $^7$Be neutrino fluxes. These reactions are the
principal terminating fusion reactions of the p-p chain. In Section~5,I give my explanation of why it
took so long for physicists to reach a consensus that new particle physics was being learned from solar
neutrino experiments.

\section{Ray Davis and Bruno Pontecorvo}
\label{sec:ray}

\begin{figure}[!ht]
\centerline{\includegraphics[width=3.5in]{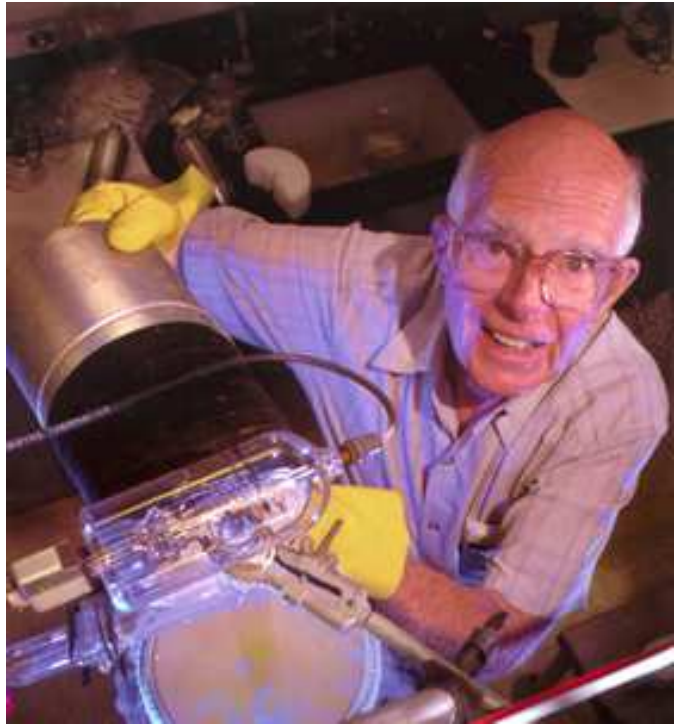}} \caption{Ray Davis preparing to pour liquid nitrogen into
a dewar on a vacuum system of the type used for gas purification and counter filling in the chlorine
experiment. The glass object in the foreground with the wire coming out that blocks Ray's left hand is an
ionization gauge used to measure the pressure in the vacuum system.\label{fig:ray}}
\end{figure}

\begin{figure}[!htb]
\centerline{\includegraphics[width=3in]{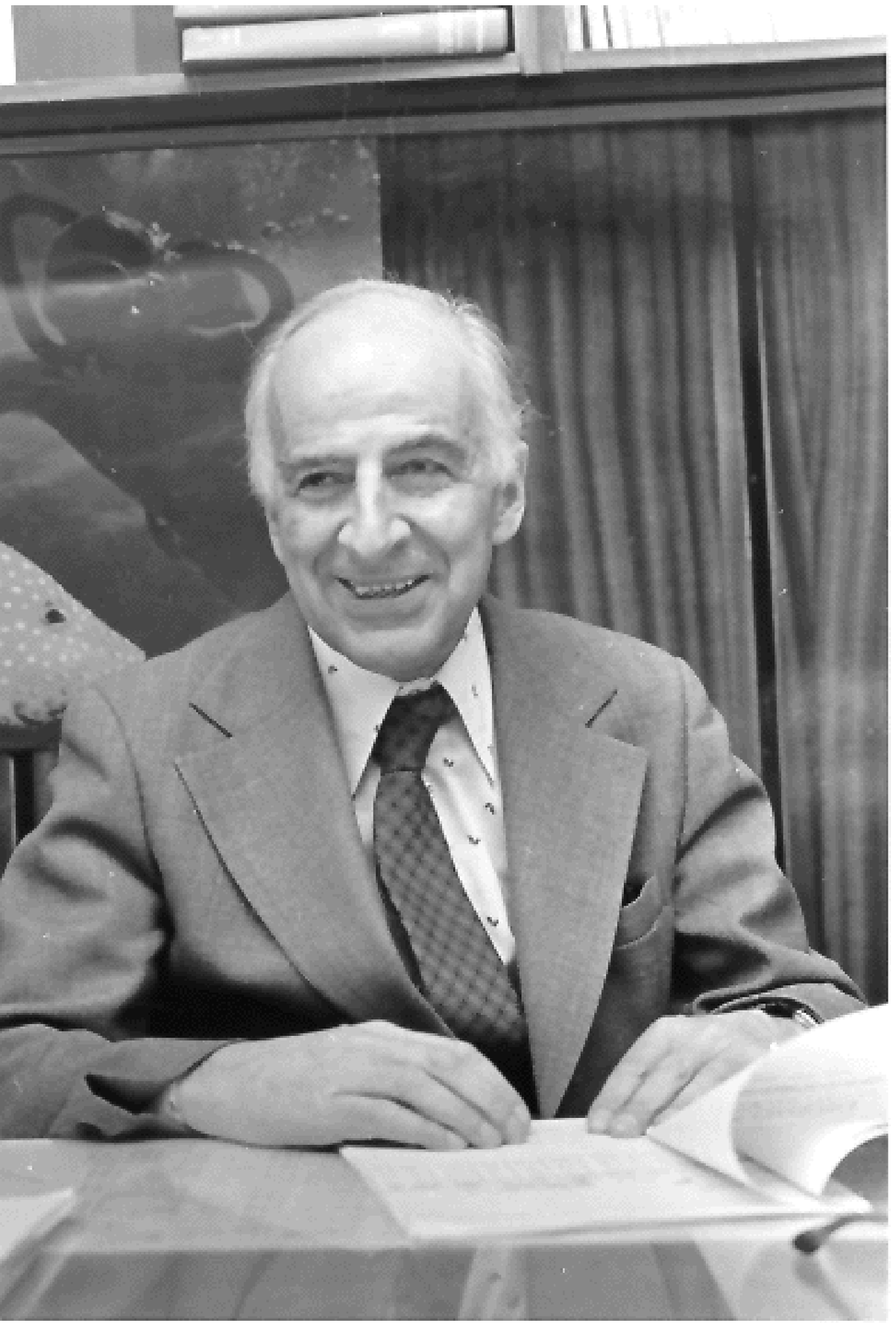}} \caption[]{Bruno Pontecorvo wrote in 1967:
`From the point of view of detection possibilities, an ideal object is the sun.' Figure courtesy of S.
Bilenky .\label{fig:pontecorvo}}
\end{figure}

Before I begin the discussion of the standard solar model, I want to begin by paying tribute to two of
the great scientists and pioneers of neutrino astrophysics, Ray Davis (Figure~\ref{fig:ray}) and Bruno
Pontecorvo (Figure~\ref{fig:pontecorvo}). Bruno first suggested using chlorine as a detector of neutrinos
in a Chalk River report written in 1946. Ray followed up on Bruno's suggestion and the careful
unpublished feasibility study of Louie Alvarez. Using with
 care and skill a chlorine detector and reactor neutrinos, Ray
showed in 1955-1958 that $\nu_e$ and $\bar\nu_e$ were different. About a decade later, Ray first detected
solar neutrinos, laying the foundation for the studies that are so widely discussed today. Bruno
recognized that solar neutrinos could potentially tell us something about particle physicists and he laid
the foundation for the modern theory of neutrino oscillations.

 The solar neutrino saga has been a community effort in which thousands of chemists,
physicists, astronomers, and engineers have contributed in crucial ways to refining the nuclear physics,
the astrophysics, and the detectors so that the subject could become a precision test of stellar
evolution and, ultimately, of weak interaction theory. But, these two wonderful scientists and wonderful
human beings started us on a marvelous road of discovery.

Ray's role in the subject, like Bruno's, has been unique. Any historical summary, even of solar models,
would be grossly incomplete if it did not emphasize the inspiration provided by Ray's experimental
vision. Although Ray never was involved in solar model calculations, and has always maintained a healthy
skepticism regarding their validity, his interest in performing a solar neutrino experiment was the
motivation for my entering and remaining in the subject. More importantly, for all of the formative years
of the ``solar neutrino problem", Ray inspired everyone  who became involved with solar neutrinos by his
conviction that valid and fundamental measurements could be made using solar neutrinos. We committed to a
subject that did not attract main stream scientists because we believed in Ray's dream of measuring the
solar neutrino flux.

In 1967, one year before the first results of Ray's chlorine solar neutrino experiment were announced,
Bruno published a prophetic paper entitled: `Neutrino Experiments and the Problem of Conservation of
Leptonic Charge' [Zh. Exp. Teor. Fiz. 53, 1717 (1967)]. In this paper, Bruno suggested many different
experiments that could test whether leptonic charge was conserved. The grandchildren of these experiments
are being discussed this summer by particle physicists at conferences held all over the world.

Bruno included a short section in his paper that he called `Oscillations and Astronomy.' In this section,
Bruno wrote: ``From the point of view of detection possibilities, an ideal object is the sun,'' What a
wonderfully contemporary statement!

Bruno, like most particle physicists of the 1960's, the 1970's, the
  1980's, and even the 1990's, did not believe astrophysical calculations could be
  reliable. He wrote in this same section on oscillations and
  astronomy: ``Unfortunately, the weight of the various thermonuclear
  reactions in the sun, and the central temperature of the sun are
  insufficiently well known in order to allow a useful comparison of
  expected and observed solar neutrinos, from the point of view of
  this article.'' [This was 30 years before the precise confirmation
  of the standard solar model by helioseismology.] To support his
  claim, Bruno referenced only his 1946 Chalk River report, which
  mentioned the sun in just two sentences.  Bruno did cite our detailed
  calculations of the solar neutrino fluxes elsewhere in his 1967
  paper, but they seem not to have affected his thinking.

Ray and I have written three articles on the history of solar neutrino research (in 1976, 1982, and 2000,
see http://www.sns.ias.edu/~jnb under the menu item Solar Neutrinos/History). It is not feasible to
present in a short talk a balanced account of all the material covered in these three articles with the
appropriate acknowledgments of the important work of so many people. Therefore, I shall just describe
some of the highlights regarding the standard solar model from a very personal view. I encourage the
listeners who are interested in a more balanced presentation to look back at the earlier articles which
provide references to critical work done by a large number of researchers.

\section{The development of the ``standard solar model" for
neutrino predictions} \label{sec:development}

I describe the development of the ``standard solar model" for neutrino predictions in five subsections,
covering the period 1962-1988 (Section~3.1), 1988-1995 (Section~3.2), 1995-1997 (Section~3.3), 1998-2002
(Section~3.4), and 2002-2003 (Section~3.5).

\subsection{1962-1988}
\label{subsec:6268}

 At the time Ray and I first
began discussing the possibility of a solar neutrino experiment,
in 1962, there were no solar model calculations of solar neutrino
fluxes. Ray, who heard about some of my work on weak interactions
from Willy Fowler, wrote and asked if I could calculate the rate
of the $^7$Be electron capture reaction  in the Sun.

After I did the calculation and submitted the paper to Physical
Review, I woke up to the obvious fact that we needed a detailed
model of the Sun (the temperature, density, and composition
profiles) in order to convert the result to a flux that Ray might
consider measuring. I moved to Willy's laboratory at  CalTech,
where there were experts in stellar modeling who were working on
stellar evolution. We used the codes of Dick Sears and Icko Iben,
and a  bit of nuclear fusion input that I provided, to calculate
the first solar model prediction of solar neutrinos in
$1962-1963$.

The result was extremely disappointing to Ray and to me, since the
event rate from neutrino capture by chlorine that I calculated
from our first flux evaluation was too small by an order of
magnitude to be measured in any chlorine detector that Ray thought
would be feasible. The situation was reversed in late 1963, when I
realized that the capture rate for $^8$B neutrinos on chlorine
would be increased by almost a factor of $20$ over my earlier
calculations because of transitions to the excited states of
argon, most importantly the super-allowed transition from the
ground state of $^{37}$Cl to the isotopic analogue state at about
$5$ MeV excitation energy in $^{37}$Ar.  This increase in the
predicted rate made the experiment appear feasible and Ray and I
wrote a joint paper for Physical Review Letters proposing a
practical chlorine experiment, a paper that was separated into two
shorter papers to meet the space requirements.

During the period $1962-1968$, the input data to the solar models
were refined in a number of important ways as the result of the
hard work of many people. The most significant changes were in the
measured laboratory rate for the $^3$He-$^3$He reaction (changed
by a factor of 3.9), in the theoretically calculated rate for the
$p-p$ reaction (changed by $7$\%), and the observed value of the
heavy element to hydrogen ratio, $Z/X$ (decreased by a factor of
$2.5$). Unfortunately, each of the individual corrections were in
a direction that decreased the predicted flux.

Ray's first measurement was reported in PRL in 1968. Our
accompanying best-estimate solar model prediction (made together
with N. A. Bahcall and G. Shaviv) was about a factor of 2.5 times
larger than Ray's upper limit. But the uncertainties in the model
predictions were, in 1968, sufficiently large that I personally
did not feel confident in concluding that the  disagreement
between prediction and measurement meant that something
fundamental was really wrong.

As it turned out, the values of the stellar interior parameters
used in $1968$ are in reasonably good agreement with the values
used today. However, the uncertainties are much better known now,
after more than three decades of intense and precise studies and
refinements by many different groups working all over the world.

The laboratory  measurement of the $^7$Be($p,\gamma$)$^8$B cross
section was a principal source of uncertainty in the $1962$
prediction, remained a principal uncertainty in $1968$, and is
still today one of the two largest uncertainties in the solar
neutrino predictions. Moreover, the best-estimate measured value
for the cross section has decreased  significantly since 1968 (see
Figure~\ref{fig:be7pcrosssection}).

As we shall see in the subsequent discussion, the only
fundamentally new element that has been introduced in the
theoretical calculations since $1968$ is the effect of element
diffusion in the sun (see 1988-1997 below).

During the period $1968-1988$, very few people worked on topics
related to solar neutrinos. There was only one solar neutrino
detector, Ray's chlorine experiment. His measurement was lower
than our prediction. I concentrated during these two long decades
on refining the predictions and, most importantly, making the
estimates of the uncertainties more formal and more robust.

We calculated the uncertainties by computing the partial
derivatives of each of the fluxes with respect to each of the
significant input parameters. In 1988, Roger Ulrich and I also did
a Monte Carlo study of the uncertainties, which made use of the
fluxes calculated from $1000$ standard solar models. For each of
the $1000$ models, the value of each input parameter was drawn
from a probability distribution that had the same mean and
variance as was assigned to that parameter.  The Monte Carlo
results confirmed the conclusions reached using the partial
derivatives. The uncertainty estimates made during this period are
the basis for the uncertainties assigned in the current neutrino
flux predictions and influence inferences regarding neutrino
parameters (like $\Delta m^2$, $\tan^2 \theta$) that are derived
from analyses that make use of the solar model predictions.

\subsection{1988-1995}
\label{subsec:8895}

In the period $1990-1994$, F. Rogers and J. Iglesias of the
Livermore National Laboratory published their detailed and
improved calculations of stellar radiative opacities and equation
of state. Now almost universally used by stellar modelers, this
fundamental work resolved a number of long standing discrepancies
between observations and predictions of stellar models.

 In the same $1988$ RMP paper in which we presented the Monte Carlo study
of the uncertainties, Roger Ulrich and I also made comparisons
between the predictions of our standard solar model--constructed
to predict solar neutrinos--and the then existing
helioseismological data on $p$-mode oscillations. The agreement
was reasonably impressive: the model predictions and the measured
frequency splittings  agreed to about $0.5$\%. But, we suspected
that there was something missing in the solar models.

During the period $1990-1995$, my colleagues and I made
successively better approximations at including element diffusion
in the solar model calculations. First, we derived an approximate
analytic description which was included in the solar models (after
some significant coding struggles) and later we made use of a
precise computer subroutine that calculated the diffusion
numerically. This work was done with S. Basu, A. Loeb, M.
Pinsonneault, and A. Thoule.

\subsection{1995-1997}
 \label{subsec:9597}
In $1995$, Steve Tomczyk and his colleagues presented the first
observations of the solar $p$-mode oscillations that included
modes that sampled well both the intermediate solar interior and
the deep interior. These observations determined precise
observational values for the sound speed over essentially the
entire solar interior.

We were in a wonderful position to make use of these precise sound
speeds. In $1995$,  Marc Pinsonneault and I had just published a
systematic study of improved solar models that incorporated the
new opacity and equation of state calculations from the Livermore
group and, most importantly, we had succeeded in including helium
and heavy element diffusion in our standard solar model.

\begin{figure}[!t]
\centerline{\includegraphics[width=3.5in]{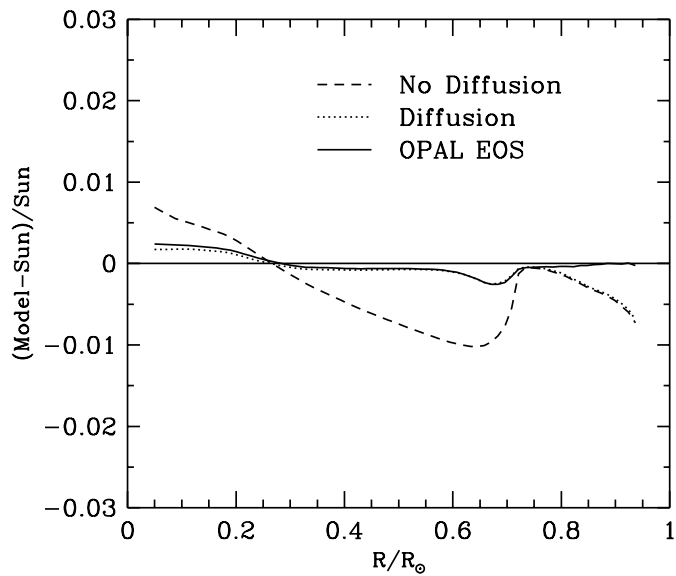}}
\caption[]{Comparison of measured and calculated
sound speeds. This figure from astro-ph/9610250, PRL {\bf 178},171
1997 compares the sound speeds calculated with the standard solar
model, BP95, with the helioseismologically determined sound
speeds. The dashed curve represents the results from a solar model
that does not include element diffusion. Much better agreement is
obtained when element diffusion is included, as indicated by the
dotted curve. The solid line represents a model in which both
element diffusion and the refined OPAL equation of state are
included. The inclusion of the refined equation of state results
in a slight improvement in the innermost region of the solar
model.\label{fig:soundspeeds}}
\end{figure}

Together with Sarbani Basu and Joergen Christensen-Dalsgaard, we
showed that the helioseismologically measured sound speeds were in
excellent agreement throughout the Sun with the values calculated
from our previously constructed standard solar model. As shown in
Figure~\ref{fig:soundspeeds}, the agreement averaged better than
$0.1$\% r.m.s. in the solar interior. We made a simple scaling
argument between accuracy in predicting sound speeds and accuracy
in predicting neutrino fluxes. The concluding sentence in the
Abstract of our PRL paper was:

``Standard solar models predict the structure of the Sun more
accurately than is required for applications involving solar
neutrinos."

This result was published in the January 1997 issue of PRL, but I had earlier presented at Neutrino '96
in Helsinki (June 1996) the same conclusion based upon somewhat less precise helioseismological data.
Since some of you were present also at Helsinki, you may be interested in the precise form of the
statement made in the printed proceedings:

``Helioseismology, as summarized in Figure 2 [a comparison of
measured and calculated sound speeds], has effectively shown that
the solar neutrino problems cannot be ascribed to errors in the
temperature profile of the Sun."

The helioseismological confirmation of the standard solar model changed for me personally the way that I
regarded `the solar neutrino problem.' I no longer felt it was necessary to soften the claim that the
origin of the `problem' was new physics not bad astronomy.

So, from the astronomical perspective, we have known for six years that new physics was required to
resolve the discrepancy between the standard predictions of the solar model and electroweak theory. Even
prior to the existence of this helioseismological evidence, it had become clear that one could not fit
the data for all the solar neutrino experiments by simply re-scaling standard predictions of neutrino
fluxes.

Why did it take so long? In Section~5, I will try to answer the question: Why were some physicists
unconvinced by the astronomical evidence that solar neutrino oscillations occurred?

\subsection{1998-2002}
\label{subsec:9802}

The SNO and Super-Kamiokande experiments have confirmed directly the calculated solar model flux of $^8$B
neutrinos, provided there is not a large component of sterile neutrinos in the incident flux.

In units of $10^6{\rm ~ cm^{-2}s^{-1}}$, the standard solar model
prediction for the flux, $\phi$, of rare $^8$B neutrinos is

\begin{equation}
\phi({\rm BP00}) ~=~ 5.05^{+1.0}_{-0.8}. \label{eq:bp00prediction}
\end{equation}
In June 2001, the SNO collaboration announced that the combined
result from their initial CC measurement and the Super-Kamiokande
$\nu-e$ scattering measurement implied a flux of $^8$B active
neutrinos equal to

\begin{equation}
\phi({\rm SNO~CC \,+\, SK}) ~=~ 5.44 \pm 0.99. \label{eq:SKplusCC}
\end{equation}
The agreement between the best-estimate calculated value given in
Eq.~(\ref{eq:bp00prediction}) and the best-estimated measured
value given in Eq.~(\ref{eq:SKplusCC}) is  $0.3\sigma$.

The recent SNO NC measurement implies an even closer agreement
between the best-estimates. Assuming an undistorted $^8$B neutrino
spectrum (a very good approximation), the SNO collaboration finds

\begin{equation}
\phi({\rm NC}) ~=~ 5.09 \pm 0.64. \label{eq:SNOnc}
\end{equation}

The agreement between the best-estimates given in
Eq.~(\ref{eq:bp00prediction}) and Eq.~(\ref{eq:SNOnc}) is
embarrassingly small, $0.03\sigma$, but obviously accidental. The
quoted errors, theoretical and experimental, are real and
relatively large.

\subsection{2002-2003:Post-KamLAND}
\label{subsec:postkamland}

Very recently, the results of the KamLAND reactor experiment (hep-ex/0212021) have led to a more precise
determination of both the $^8$B and (after this meeting was held) the $p-p$ neutrino fluxes.

We have recently completed a global analysis of all the available solar and reactor data (see
hep-ph/0212147 and hep-ph/0305159), including especially the KamLAND measurements.  The agreement with
the standard solar model predictions is good. When expressed in terms of the standard solar model (BP00)
predicted neutrino flux, the experimentally determined flux of $^8$B solar neutrinos is

\begin{equation}
\phi(^8{\rm B})~=~ 1.00 \pm 0.04 . \label{eq:b8postkamland}
\end{equation}
The experimentally-determined flux of $p-p$ solar neutrinos, expressed in terms of the BP00 predicted
flux, is

\begin{equation}
\phi({\rm p-p})~=~ 1.01 \pm 0.02.
 \label{eq:pppostkamland}
\end{equation}

We shall now discuss uncertainties in the predictions of the solar
neutrino fluxes.

\section{Uncertainties in the solar model predictions}
\label{sec:uncertainties} I will  begin the discussion of uncertainties with a brief introduction in
Section~4.1 that emphasizes the importance of robust and well-defined estimates of the errors. Then I
will describe in Section~4.2 the most important sources of uncertainties in the contemporary predictions.

\subsection{Skepticism}
\label{subsec:skepticism}

>From the very beginning of solar neutrino research, the uncertainties in the solar model predictions have
been a central issue. We could not learn things about the Sun or about neutrinos using solar neutrino
experiments unless we could demonstrate that the uncertainties in the predictions were small and robustly
calculated. If, as many physicists initially believed, the astronomical predictions were not
quantitatively reliable, then there was no real ``solar neutrino problem."

Bruno Pontecorvo, in his prophetic paper ``Neutrino Experiments and the Problem of Conservation of Lepton
Charge", Soviet Physics JETP, 26, 984 (1968), expressed the view that the uncertainties in the solar
model calculations were so large as to prevent a useful comparison with solar neutrino experiments, Here
is what Bruno said:

``From the point of view of detection possibilities, an ideal
object is the sun... Unfortunately, the weight of the various
thermonuclear reactions in the sun, and the central temperature of
the sun, are insufficiently well known in order to allow a useful
comparison of expected and observed solar neutrinos, from the
point of view of this article."

This comment by Bruno Pontecorvo is indicative of the skepticism about solar model predictions that
existed among many physicists. In an effort to remove this skepticism, I spent much of the 34 years from
1968 to 2002 refining the predictions of the solar neutrino fluxes and providing increasingly more robust
estimates of the uncertainties in the predictions.

 I want to summarize for you now the current best estimates for the uncertainties in the solar neutrino
predictions.

\subsection{Currently estimated uncertainties in predicted neutrino fluxes}
\label{subsec:currentuncertainties}

 I will first present in
Section~4.2.1
 the current values for the
total and the partial uncertainties in the flux predictions. Then I will describe in Section~4.2.2 and
Section~4.2.3, respectively, the very different histories for the determination of the cross sections for
the $^7$Be(p,$\gamma$)$^8$B reaction and the $^{37}$Cl($\nu$, $e^-$)$^{37}$Ar.

\subsubsection{Total and fractional uncertainties}
\label{subsubsec:totalpartial}

\begin{figure}
\centerline{\includegraphics[width=3.5in,angle=270]{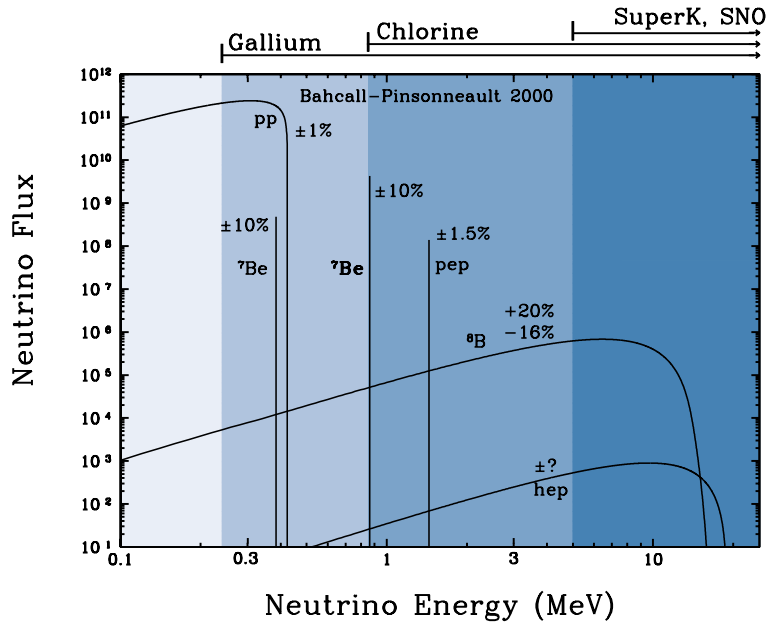}}
\caption[]{Solar neutrino spectrum with currently
estimated uncertainties. \label{fig:snspectrum}}
\end{figure}

 Figure~\ref{fig:snspectrum} shows
the calculated values for the principal (p-p)  solar neutrino
fluxes and their estimated uncertainties. The p-p and pep neutrino
fluxes are predicted with a calculated uncertainty of only $\pm
1$\% and $\pm 1.5$\%, respectively. The $^7$Be neutrino flux is
predicted  with an uncertainty of $\pm 10$\% and the important
$^8$B neutrino flux, which is measured by Super-Kamiokande and
SNO,  is predicted with an error of about $20$\%. The fluxes from
CNO reactions, especially $^{13}$N and $^{15}$O neutrino fluxes,
are predicted with less precision than the fluxes from the p-p
reactions. I have not shown the CNO fluxes in
Figure~\ref{fig:snspectrum} since these fluxes are not expected to
play a discernible role in any of the planned or in progress solar
neutrino experiments.

\begin{table}[!htb]
\centering \caption[]{Fractional uncertainties in the Predicted
$^8$B and $^7$Be  Solar Neutrino Fluxes (BP00). The table presents
the fractional uncertainties in the calculated $^8$B and $^7$Be
neutrino fluxes, due to the different factors listed in the column
labeled Source. The first four rows refer to the low energy cross
section factors for different fusion reactions. The last four rows
refer to the heavy element to hydrogen ratio, $Z/X$ (Composition),
the radiative opacity, a multiplicative constant in the expression
for the diffusion rate of heavy elements and helium, and the total
solar optical luminosity. \label{tab:uncertainties}}
\begin{tabular}{ccc}
\hline\hline
\noalign{\smallskip}
Source& $^8$B&$^7$Be\\
\noalign{\smallskip}
\hline
\noalign{\smallskip}
p-p&0.04&0.02\\
$^3$He+$^3$He&0.02&0.02\\
$^3$He+$^4$He&0.08&0.08\\
p + $^7$Be&$^{+0.14}_{-0.07}$&0.00\\
Composition&0.08&0.03\\
Opacity&0.05&0.03\\
Diffusion&0.04&0.02\\
Luminosity&0.03&0.01\\
\noalign{\smallskip}
\hline\hline
\end{tabular}
\end{table}

Table~\ref{tab:uncertainties} shows how much each of the principal
sources of uncertainty contribute to the total present-day
uncertainty in the calculation of the $^8$B and $^7$Be solar
neutrino fluxes. The largest uncertainty in the prediction of the
$^8$B neutrino flux is caused by the estimated error in the
laboratory measurement of the low energy cross section for the
$^7$Be(p,$\gamma$)$^8$B reaction (This statement was also true in
1962, 1964, 1968, ....). The largest uncertainty in the prediction
of the $^7$Be neutrino flux is due to the quoted error in the
measurement of the low energy rate for the $^3$He + $^4$He
reaction.  In addition, there are a number of other sources of
uncertainty, all of which contribute more or less comparably to
the total uncertainty in the prediction of the $^7$Be and the
$^8$B neutrino fluxes.

\begin{figure}[!t]
\centerline{\includegraphics[width=4in,angle=270]{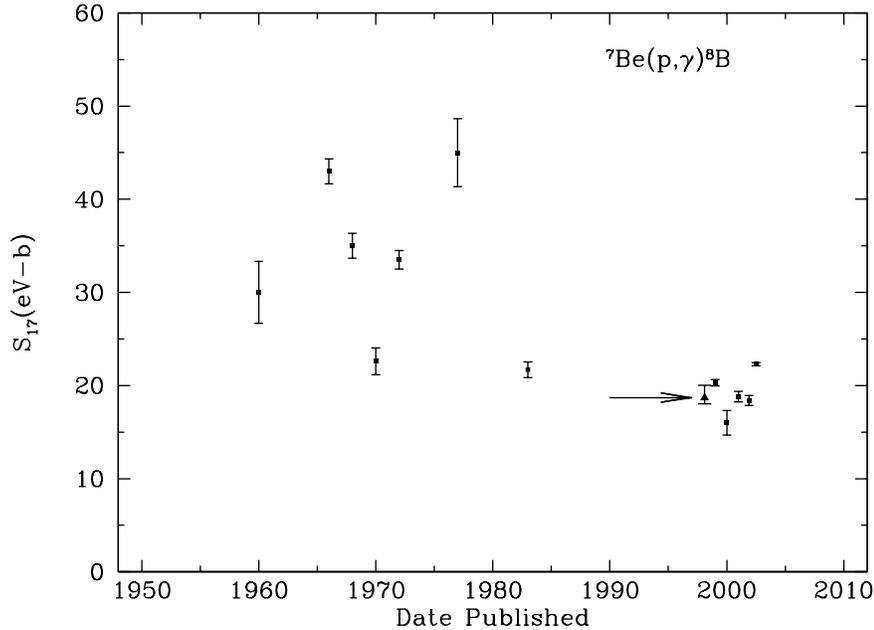}} \caption[]{$^7$Be(p,$\gamma$)$^8$B.
The figure shows the measured values as a function of date of publication for the low-energy cross
section factor for the $^7$Be(p,$\gamma$)$^8$B reaction. The arrow points to the currently standard
value, recommended by Adelberger et al., that is used in the BP00 calculations. Recent precision
measurements yield values slightly larger than the value recommended by Adelberger et al. but are
consistent with the earlier estimate [see Phys. Rev. Lett. 90 (2003) 022501 and Table~2 of
nucl-ex/0212011.] \label{fig:be7pcrosssection}}
\end{figure}

\subsubsection{The saga of the $^7$Be(p,$\gamma$)$^8$B cross section}
\label{subsubsec:saga}

 Figure~\ref{fig:be7pcrosssection} shows, as a
function of the date of publication,  measured values for the low energy cross section of the crucial
reaction $^7$Be(p,$\gamma$)$^8$B. [Some very recent measurements are included in the reference
nucl-ex/0212011.] I have only shown here the direct measurements of this reaction; there are also
indirect measurements that yield similar results.

 The
encouraging aspect of Figure~\ref{fig:be7pcrosssection} is that
the huge uncertainty that existed between 1960 and 1980, of order
a factor of two, has been much reduced in the following two
decades. In the BP00 calculations, we adopted as the best-estimate
the Adelberger et al. [RMP, 70, 1265 (1998), astro-ph/9805121]
consensus value for the cross section factor of the
$^7$Be(p,$\gamma$)$^8$B reaction, $S_{17}(0) = 19 (1 +
^{+0.14}_{-0.07})$ eV-b (the $1\sigma$ error given here is
one-third the  Adelberger et al. $3\sigma$ estimate). This value
is indicated in the figure by arrow next to ``Standard."

Several refined experiments are in progress or are planned to
measure more accurately the low energy cross section factor for
the $^7$Be(p,$\gamma$)$^8$B reaction or the
p($^7$Be,$\gamma$)$^8$B reaction. Also, there are a number of
related reactions that are being studied in order to give somewhat
more indirect information about the low energy cross section. The
goal of all these experiments is to reduce the combined systematic
and statistical errors to below 5\%, so that $S_{17}(0)$ is no
longer a dominant source of uncertainty in the prediction of the
$^8$B solar neutrino flux (cf. Table~\ref{tab:uncertainties}
above).

To the best of my knowledge, the preliminary data from all of the existing experiments are consistent
with the currently standard value of $S_{17}(0)$ quoted above and with the best-estimate recommended in
reference nucl-ex/0212011. In order to avoid the confusion that would be created by introducing numbers
in the literature that are changed frequently, I prefer not to revise the ``standard" estimate of
$S_{17}(0)$ (and the $^8$B solar neutrino flux)  until the in-progress experiments on
$^7$Be(p,$\gamma$)$^8$B and related reactions are completed.

\subsubsection{The $^{37}$Cl($\nu_e$,
$e^-$)$^{37}$Ar cross section} \label{subsubsec:nucrosssection}

\begin{figure}[!t]
\centerline{\includegraphics[width=4in,angle=270]{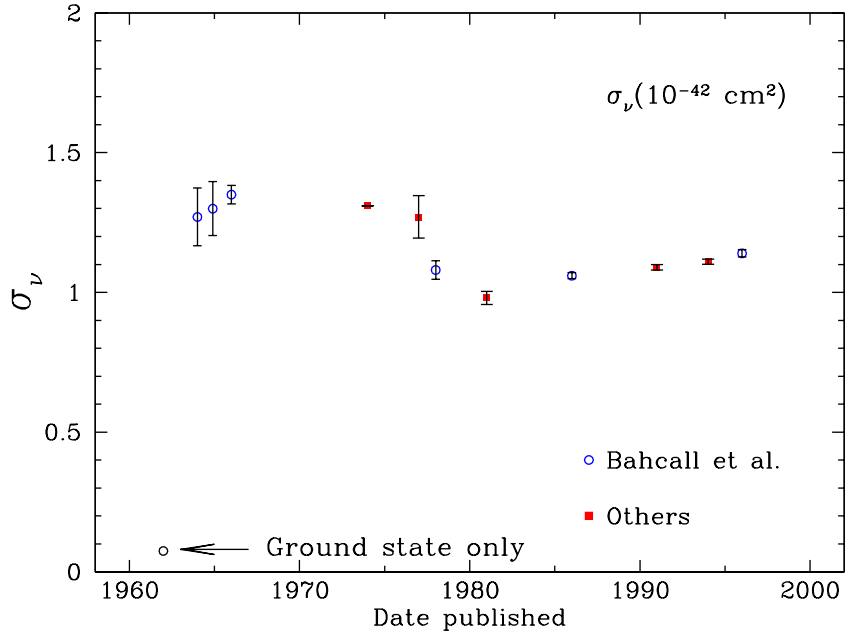}}
\caption[]{$^{37}$Cl($\nu_e$,$e^-$)$^{37}$Ar. The
figure shows, for an undistorted $^8$B solar neutrino spectrum,
the calculated values for the cross section
$^{37}$Cl($\nu_e$,$e^-$)$^{37}$Ar as a function of date of
publication.  \label{fig:nucrosssection}}
\end{figure}

 In the early days of solar neutrino
astronomy, the cross sections for neutrino absorption by chlorine,
$^{37}$Cl($\nu_e$, $e^-$)$^{37}$Ar, were an important source of
uncertainty. For comparison with
Figure~\ref{fig:be7pcrosssection}, I show in
Figure~\ref{fig:nucrosssection} the calculated values of the
absorption cross section for $^8$B neutrinos incident on
$^{37}$Cl. The first calculation I made (in 1962) was too small,
because I did not consider transitions to excited states. The
calculation I made in 1964 was quickly confirmed by measurements
made (by Poskanzer et al.) on the predicted decay: $^{37}$Ca
$\rightarrow$ $^{37}$K + $e^+$ + $\nu_e$, which is the isotopic
analogue of the neutrino capture reaction. A series of subsequent
refined measurements and calculations reduced the estimated error
in the neutrino cross section to where it is no longer one of the
largest sources of uncertainty in the calculation of the predicted
capture rate in the chlorine solar neutrino experiment (although
the uncertainty still plays some role in the global determination
of solar neutrino oscillation parameters).

\subsection{Using p-p and $^7$Be neutrinos to probe details of solar fusion}
\label{subsec:probe}

 Is there any way of probing the solar
interior and determining experimentally which terminating reaction
of the p-p chain, $^3$He-$^3$He or $^3$He-$^4$He, is faster in the
solar interior and by how much? Yes, there is a way. Solar
neutrino experiments can do just that.

 The ratio R of
the rate of $^3$He-$^3$He reactions to the rate of $^3$He-$^4$He
reactions averaged over the Sun can be expressed in terms of the
p-p and $^7$Be neutrino fluxes by the following simple
relation\footnote{More precisely, $\phi(^7{\rm Be})$ should be
replaced by the sum of the $^7$Be and $^8$B neutrino fluxes in the
denominator of Eq.~(\ref{eq:defnR}).}:

\begin{equation}
R ~\equiv~\frac{<^3{\rm He} + ^4{\rm He}>}{<^3{\rm He} + ^3{\rm
He}>} ~=~\frac{2\phi(^7{\rm Be})}{\phi({\rm pp})~-~\phi(^7{\rm
Be})}. \label{eq:defnR}
\end{equation}

The standard solar model predicts $R = 0.174$.  One of the reasons
why it is so important to measure accurately the total p-p and
$^7$Be neutrino flux is in order to test this detailed prediction
of standard solar models.  The value of R reflects the competition
between the two primary ways of terminating the p-p chain and
hence is a critical probe of solar fusion.

\section{Why did it take so long?}
\label{sec:why}

In the introduction to this talk, I said that I would address the
question of why it took so long, about 35 years, to convince many
physicists that solar neutrino research was revealing something
new about neutrinos. I will now do my best to explain why the
process from discovery to consensus required more than three
decades.

In the early years,  after the very rapid progress between 1964 to 1968, there were  many things that had
to be studied very carefully to see if there could be something important that had been left out of the
standard solar models. The values of all of the (large number of) important input parameters were
remeasured or recalculated more accurately, a variety of imaginative ``non-standard solar models" were
examined critically, and possible instabilities in the solar interior were investigated. It took about 20
years, 1968-1988, for the collective efforts of many nuclear physicists, atomic physicists, astronomers,
and astrophysicists to provide a thoroughly explored basis for the standard model calculations that
allowed robust estimates of the uncertainties in the solar model predictions. Even after this long
struggle with details was mostly complete, it was still necessary to develop codes that could include the
refinement of element diffusion (which took until 1995). And, presumably, there are still today even
further refinements that are appropriate and necessary to make to obtain a still more accurate
description of the region in which solar fusion takes place.

My impression is that nearly all particle physicists remained
blissfully unaware of, or indifferent to, the decades of efforts
to make the solar neutrino predictions more robust. Why? Why did
many (but not all) particle physicists not take the ``solar
neutrino problem" seriously?

I think that there were three reasons it took so long for particle physicists to acknowledge that new
physics was being revealed in solar neutrino research. First, the Sun is an unfamiliar accelerator.
Particle physicists, and most other physicists too, were skeptical of what astronomers and
astrophysicists could learn about an environment that they could neither visit nor manipulate. These
physicists often had only a newspaper-level understanding of the observational phenomena that stellar
models reproduced and the constraints they met. Second, physicists who heard talks on solar neutrinos,
were most impressed by the fact that the $^8$B solar neutrino flux depended on the 25th power of the
central temperature, $\phi(^8{\rm B}) \propto T^{25}$. This dependence seemed to many physicists too
sensitive to allow an accurate prediction (an objection which was answered experimentally only by the
helioseismological measurements in 1995 and their successful comparison in 1996-1997 with standard solar
model predictions, see Section~3.3.) Third, the simplest interpretation of the discrepancy between
observed and predicted solar neutrino event rates, vacuum neutrino oscillations (proposed by Bruno
Pontecorvo), suggested large mixing angles for the neutrinos. It was widely (but not universally) agreed
among particle theorists that mixing angles in the lepton sector would be small in analogy with the
mixing angles in the quark sector. The most popular view of particle theorists over most of the history
of solar neutrino research has been that since quarks and leptons are probably in the same multiplets,
they should have mixing angles of comparable size. This objection to new solar neutrino physics was
removed only when Mikheyev and Smirnov built upon the earlier work of Wolfenstein to describe the magic
of the MSW effect. Ironically, the small mixing angle (SMA) MSW solution persuaded a significant number
of  physicists that there might be new physics being revealed by solar neutrino experiments, although
today we know that only large mixing angles solutions are good fits to all the available solar and
reactor neutrino data.

I think that the spirit with which many particle physicists
regarded solar neutrino research is best expressed by a quotation
from the introduction of a 1990 paper written by H. Georgi and M.
Luke [Nucl. Phys. B, 347, 1 (1990)]. They began their article as
follows:

``Most likely, the solar neutrino problem has nothing to do with
particle physics. It is a great triumph that astrophysicists are
able to predict the number of $^8$B neutrinos to within a factor
of 2 or 3..."

This writeup is based on my talks at PIC03, at Neutrino2002 (Munich) and at Venice Telescopes 2003
(organized by Milla Baldo Ceolin). This work was partially supported by an NSF grant No. PHY0070928.

\end{document}